\PassOptionsToPackage{table,dvipsnames}{xcolor}
\documentclass[acmsmall]{acmart}
\AtBeginDocument{%
  }

\setcopyright{acmlicensed}
\copyrightyear{2018}
\acmYear{2018}
\acmDOI{XXXXXXX.XXXXXXX}
\acmConference[xx]{xx}{June 03--05,
  2018}{Woodstock, NY}
\acmISBN{978-1-4503-XXXX-X/2018/06}
\usepackage[utf8]{inputenc}
\usepackage{xcolor}     
\usepackage{array}      
\usepackage{colortbl}   
\usepackage{booktabs}
\usepackage{graphicx}
\usepackage{subcaption}
\usepackage{float}
\usepackage{xcolor}
\definecolor{mygreen}{RGB}{217, 234, 211} 
\definecolor{myyellow}{RGB}{255, 242, 204} 
\definecolor{myred}{RGB}{244, 204, 204} 

\begin{document}

\title[Designing 1-Minute Micro Health Interventions]{Micro-Health Interventions: Exploring Design Strategies for 1-Minute
Interventions as a Gateway to Healthy Habits}

\author{Zahra Hassanzadeh}
\email{hassanzadeh@cs.toronto.edu}
\affiliation{%
  \institution{University of Toronto}
  \city{Toronto}
  \state{Ontario}
  \country{Canada}
}

\author{David Haag}
\affiliation{%
  \institution{ Ludwig Boltzmann Institute for Digital Health and Prevention}
  \city{Salzburg}
  \country{Austria}}
\email{david.haag@dhp.lbg.ac}

\author{Lydia Chilton}
\affiliation{%
  \institution{ Columbia University}
  \city{New York}
  \country{United States}
}

\author{Jan Smeddinck}
\affiliation{%
 \institution{Ludwig Boltzmann Institute for Digital Health and Prevention}
 \city{Salzburg}
 \country{Austria}}

\author{Norman Farb}
\affiliation{%
  \institution{University of Toronto}
  \city{Toronto}
  \country{Canada}}

\author{Joseph Jay Williams}
\affiliation{%
  \institution{University of Toronto}
  \city{Toronto}
  \country{Canada}}

\renewcommand{\shortauthors}{Hassanzadeh et al.}

\begin{abstract}
 One-minute behavior change interventions might seem too brief to matter. Could something so short really help people build healthier routines? This work explores this question through two studies examining how ultra-brief prompts might encourage meaningful actions in daily life. In a formative study, we explored how participants engaged with one-minute prompts across four domains: physical activity, eating, screen use, and mental well-being. This revealed two common design approaches: Immediate Action prompts (simple, directive tasks) and Reflection-First prompts (self-awareness before action). We then conducted a 14-day, within-subjects study comparing these two flows with 28 participants. Surprisingly, most participants did not notice differences in structure—but responded positively when prompts felt timely, relevant, or emotionally supportive. Engagement was not shaped by flow type, but by content fit, tone, and momentary readiness. Participants also co-designed messages, favoring those with step-by-step guidance, personal meaning, or sensory detail. These results suggest that one-minute interventions, while easily dismissed, may serve as meaningful gateways into healthier routines—if designed to feel helpful in the moment.
\end{abstract}




\maketitle

\section{Introduction}
Healthy habits are defined in the behavior change and health psychology literature as regular, automatic behaviors that contribute positively to physical, emotional, or mental well-being and are maintained over time with minimal conscious effort once established \cite{fogg2020, clear2018,Jenkins2021}. Examples include drinking water instead of sugary beverages, taking short walking breaks during the day, practicing one-minute mindfulness exercises, limiting screen time before bed, or preparing balanced meals. Building healthy habits is important for improving well-being and reducing the risk of chronic illnesses such as obesity, diabetes, cardiovascular disease, and depression. Small, sustainable actions—like choosing water over sugary drinks, taking brief walking breaks, practicing one-minute mindfulness exercises, or limiting evening screen time—have been linked to significant improvements in physical and mental health outcomes over time \cite{fogg2020, Jenkins2021, kabat1990}. Behavior change literature consistently emphasizes that even micro-behaviors can accumulate into meaningful health benefits, improving resilience against chronic conditions. However, a persistent psychological barrier to behavior change is the widespread belief that strong motivation must precede action. Early behavior change theories, such as the Transtheoretical Model of Change \cite{Prochaska1983} and Bandura’s theory of self-efficacy \cite{Bandura1997}, framed readiness and perceived motivation as prerequisites for initiating behavior. More recent models, such as Behavioral Activation Therapy \cite{jacobson2001} and Fogg’s Behavior Model \cite{fogg2009}, challenge this assumption, suggesting that small, intentional actions themselves can spark motivation and support sustained behavior change.

One key insight from the behavior change literature is that starting small is critical for building sustainable habits. Fogg’s Behavior Model \cite{fogg2009} provides a foundational framework for understanding how tiny interventions can drive behavior change. According to the model, behavior (B) occurs when motivation (M), ability (A), and a prompt (P) converge at the same moment (B = MAP). This means that even when motivation is low, behavior can still occur if the action is sufficiently simple and the prompt is delivered at the right time. By minimizing the effort required and embedding easy actions into daily routines, individuals can build positive momentum. Carefully timed prompts serve as critical triggers, ensuring that new habits are initiated without requiring extensive willpower or planning. This "tiny habit" \cite{fogg2020} approach fosters early success and gradually reinforces long-term habit formation.

Recent work in behavioral science and HCI has explored the potential of ultra-short interventions—defined as actions or prompts lasting one to two minutes—to support everyday health and habit change. In exercise science, approaches such as “exercise snacks” and vigorous intermittent lifestyle physical activity (VILPA) has shown that short bursts of intense movement can lead to health benefits \cite{Jenkins2021}, lowering the threshold for engagement. Similarly, in digital health, micro-interventions—like short messages promoting hydration, mindfulness, or screen-time reduction—have shown early promise when they are timely and contextually aligned with users’ needs \cite{Jenkins2021, Baumel2020}. Yet sustaining engagement remains a persistent challenge, with prior studies documenting sharp drop-offs in adherence after the novelty fades \cite{Waller2009, Baumel2019}. One proposed avenue for fostering more lasting effects is the use of reflective prompts. By encouraging brief moments of metacognition—asking users to pause, assess their current state, or connect the prompt to personal goals—reflection-first interventions may deepen internal motivation and support longer-term habit development. However, it remains unclear whether these more reflective designs are feasible in everyday contexts or if they actually improve engagement compared to more direct, action-oriented prompts.
At first glance, the idea of a one-minute behavior change intervention might seem unconvincing. Can such a brief interaction truly support habit formation—or is it too minimal to matter? Despite growing interest in ultra-short digital prompts, we know surprisingly little about how people experience them in real life, or what makes them feel worth doing. Motivated by these open questions, we conducted a formative study to explore when and how one-minute interventions might encourage meaningful engagement. This led us to examine two core questions:
\begin{itemize}
    \item RQ1: How can we encourage individuals to engage in one-minute behavior change interventions?
    \item RQ2: How should one-minute behavior change interventions be phrased to encourage engagement?
\end{itemize}

The content areas we selected —healthy eating, physical activity, mindful screen time, and mental well-being— were chosen because they represent fundamental domains where small, consistent behavioral changes can lead to significant improvements in overall health and quality of life. For instance, research has shown that mindful eating habits reduce the risk of chronic conditions such as obesity and diabetes  \cite{cordeiro2015, morrison2022}, while interventions targeting screen time have been linked to improved focus, reduced stress, and better sleep \cite{brandtzaeg2018, mark2018}. Similarly, brief interventions for mental well-being, such as mindfulness practices, have demonstrated a positive impact on emotional regulation and stress reduction \cite{deci2020, kabat1990}, and short bursts of physical activity have been associated with increased energy and mood \cite{rooksby2014, smith2023}.

\begin{figure}[htbp]
  \centering
  \begin{subfigure}{0.48\textwidth}
    \includegraphics[height=10cm, width=\linewidth]{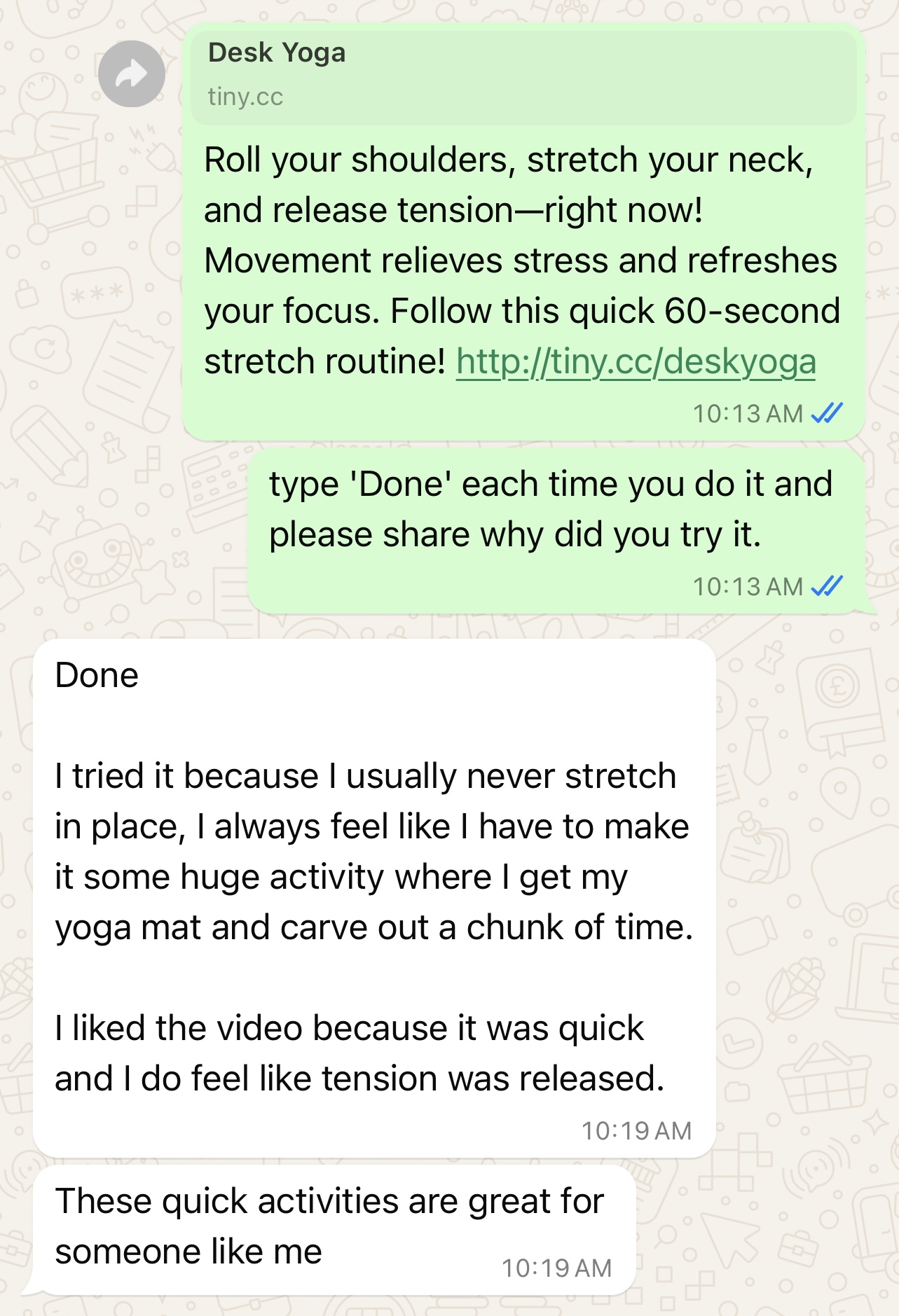}
   \caption{}
  \end{subfigure}
  \hfill
  \begin{subfigure}{0.48\textwidth}
    \includegraphics[height=8cm, width=\linewidth]{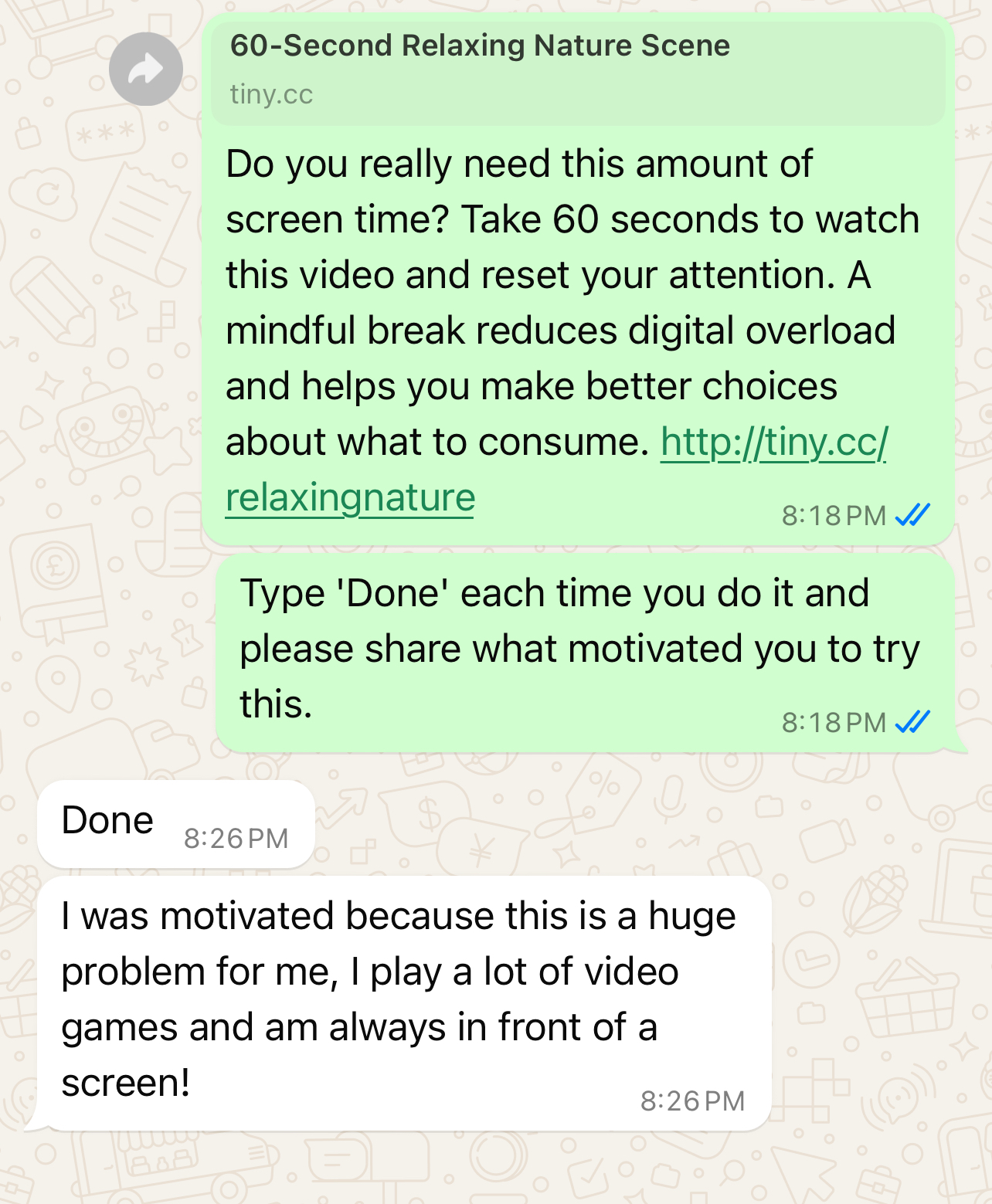}
    \caption{}
  \end{subfigure}
  \caption{Examples of prompts sent to participants: (a) Immediate Action flow (b) Reflection-First flow}
   \Description{Examples of prompts sent to participants, showing Immediate Action flow and Reflection-First flow.}
  \label{fig:sidebyside}
\end{figure}

Using WhatsApp as the delivery platform, we sent participants brief prompts designed to be completed in about one minute. These prompts encouraged immediate healthy actions, such as stretching, breathing exercises, or mindful eating choices. Fig.\ref{fig:sidebyside} shows examples of the two prompt styles used during the study: Immediate Action prompts and Reflection-First prompts that invited users to think before acting.

From the formative study, we observed that participants engaged with prompts in two distinct ways. Some gravitated toward direct, actionable tasks that required little deliberation, while others responded more strongly to prompts that invited a brief pause and reflection before acting. These contrasting interaction patterns led us to consider whether the structure of the prompt —not just its content— might meaningfully shape engagement. 

Drawing from behavior change theory, one might hypothesize that reflection-first prompts —by encouraging momentary self-awareness— could enhance internal motivation or self-efficacy and potentially lead to more meaningful follow-through \cite{ryan2000, Bandura1997}. In contrast, action-first prompts may lower the activation threshold and support quick compliance \cite{fogg2009}, but might engage users more superficially. While our study did not directly measure constructs such as self-efficacy or intrinsic motivation, we used engagement metrics —task completion rates, interaction frequency, and response depth— as behavioral proxies to examine how structural differences in flow might affect user experience.
This led us to refine our inquiry and pose a third research question:
\begin{itemize}
    \item RQ3: How do different interaction flows (Immediate Action versus Reflection-First) influence user engagement with one-minute behavior change prompts?
\end{itemize}

This paper makes the following contributions:
\begin{itemize}
    \item It examines engagement dynamics over time in the context of ultra-short, one-minute healthy habit interventions.
    \item It presents a formative design probe that identified two emergent engagement styles: Immediate Action and Reflection-First flows.
    \item It operationalizes and evaluates these styles in a structured study, offering insights into how prompt framing affects engagement.
    \item It contributes to understanding the motivational and contextual factors that drive or hinder engagement with ultra-brief digital interventions.
\end{itemize}

\section{Related Work}
Our review of related work begins with behavioral change models—including Fogg’s Behavior Model and Self-Determination Theory—as a foundation for understanding how digital prompts can support motivation, ability, and timely action. We then examine text message interventions and micro-intervention formats, with a focus on ultra-brief, one-minute prompts delivered in everyday contexts. This is followed by work on conversational agent-based interventions, including Wizard of Oz studies that simulate human-like support in lightweight, scalable ways. We also highlight the role of co-creation and participatory design in making interventions personally meaningful and contextually appropriate. Finally, we address the persistent challenges of sustaining engagement over time, including both motivational drivers and common barriers to continued use.

\subsection{Behavioral Change Models and Ultra-Short Digital Interventions
}
Our work draws on established psychological models that explain how behavioral habits can be formed and sustained. Fogg’s Behavior Model (FBM) \cite{fogg2009} highlights that behavior happens when motivation, ability, and a prompt all come together at the same time. FBM has been widely used in persuasive technology and digital health design. For example, Park et al. applied FBM in a mobile system for diabetes self-management by tailoring prompts to users’ behavioral states \cite{Park2020}, while Agha et al. used the model to study predictors of health behavior adoption (e.g., vaccination, contraception) across low-resource settings \cite{Agha2022}. Meekers et al. adapted a contraceptive program during COVID-19 using FBM to maintain usage rates through well-timed prompts \cite{Meekers2020}, and a systematic review by Orji et al. found FBM to be a dominant model in persuasive eHealth design \cite{Orji2024}. These applications underscore FBM’s relevance in structuring low-effort, prompt-driven interventions that fit seamlessly into daily life. While our study builds on models like FBM to understand how prompts support momentary action, it also highlights aspects that fall outside its traditional focus—such as reflection, personalization, and user-defined meaning. In doing so, our work complements critiques of persuasive system design that point to its tendency to narrowly frame behavior change as compliance or optimization, overlooking broader values and lived experience \cite{Brynjarsdottir2012}. We consider not only behavioral triggers, but also how ultra-brief interventions can support agency, relevance, and emotional resonance in everyday life.

Self-Determination Theory (SDT) emphasizes the role of intrinsic motivation in sustaining behavior change. It posits that fulfilling basic psychological needs—autonomy, competence, and relatedness—enhances internal motivation. In digital health, SDT has informed the design of mobile interventions that support user agency and emotional connectedness. For instance, SMS-based programs that encourage self-set goals and personalized reminders have been shown to improve engagement and satisfaction with physical activity routines \cite{Pope2017}.

Recent studies have shown that even ultra-short interventions can lead to meaningful outcomes. For example, physical health research demonstrates that short bursts of physical activity, such as “exercise snacks” or vigorous intermittent lifestyle physical activity (VILPA), can significantly improve cardiovascular fitness and metabolic outcomes \cite{rooksby2014, jacobson2001}. In the digital domain, micro-interventions—brief, lightweight digital prompts—have been used to promote small behavioral shifts without requiring sustained cognitive effort \cite{Jenkins2021, NahumShani2020}. However, while these approaches lower the threshold for action, questions remain about how different prompt framings impact real-world engagement patterns. 

\subsection{Messaging Based and Conversational Interventions for Digital Behavior Change}
Messaging platforms—especially SMS, WhatsApp, and other lightweight communication tools—have become widely adopted in digital health research for their scalability, accessibility, and integration into daily life. These platforms are particularly effective for reaching populations who may face barriers to engaging with more resource-intensive digital tools, such as mobile apps or web-based programs \cite{Bhattacharjee2022-2, Muench2017}. Text-based interventions have been applied to promote various health-related behaviors. Text-based interventions have also been applied to support mental well-being, healthy eating, and reduction in screen time. For instance, interventions like Cope Notes and HealthySMS have delivered daily supportive messages rooted in cognitive behavioral therapy and positive psychology, showing improvements in anxiety, mood, and emotional regulation \cite{Loper2020, Aguilera2017}. Programs such as BRAVE have used culturally grounded text messaging to promote mental wellness among Indigenous youth populations \cite{Rushing2011}. In the domain of healthy eating, interventions like Intuitive Eating SMS and Txt4HappyKids have used personalized messages to promote mindful eating and parental guidance around nutrition, with positive user perceptions and emerging behavioral effects \cite{Liszewski2023, Hall2015}. Similarly, screen time reduction interventions have employed text messaging to prompt healthier digital habits, particularly among children and families, with feasibility and acceptability demonstrated in both school-based and family-centered studies \cite{Wu2016, Adams2014}. Text messaging has also demonstrated effectiveness in additional domains such as smoking cessation \cite{Villanti2022, liao2016}, improving sleep hygiene \cite{Gipson2019}, supporting medication adherence \cite{Glasner2022}, and increasing vaccination uptake \cite{Buttenheim2022, Mehta2022}, among various other health-related behaviors. Many text messaging interventions draw on established frameworks from clinical psychology to structure and deliver supportive content. This includes approaches such as cognitive behavioral therapy (CBT) \cite{Wilson2019}, dialectical behavior therapy (DBT) \cite{Linehan2014}, acceptance and commitment therapy (ACT) \cite{Hayes2004}, and motivational interviewing (MI) \cite{Hettema2005}. In addition, some interventions have incorporated principles from positive psychology (e.g., gratitude and strengths-based messaging) \cite{Arps2022}, mindfulness-based cognitive therapy (MBCT)\cite{Zernicke2014}, and behavioral activation to encourage emotional regulation, resilience, and goal-directed activity \cite{Watson2017}.

Recent work in human-computer interaction (HCI) and health informatics emphasizes the importance of personalization, timing, and contextual awareness in message design \cite{fogg2020, deci2020, kabat1994}. For example, systems like TableChat have shown that embedding behavioral prompts into familiar family routines can promote sustained engagement by leveraging social support \cite{mark2018}. Just-in-Time Adaptive Interventions (JITAIs) further this goal by using real-time behavioral and contextual data to deliver messages when users are most likely to respond \cite{deci2020, kabat1994}. However, while formative investigations into scalable and deep personalization based on generative AI are emerging \cite{Haag2025}, JITAIs  remain complex to deploy. This has led to growing interest in ultra-brief, micro-intervention formats—such as one-minute prompts—that reduce cognitive burden while still maintaining behavioral salience.

Complementing these approaches, conversational interventions have increasingly used Wizard-of-Oz (WoZ) methodologies to simulate automated systems and evaluate message flow, tone, and personalization prior to full automation \cite{kabat1990, rolls2014}. Studies show that the balance between system-driven and user-driven interactions can affect engagement \cite{brandtzaeg2018}, and recent work has explored the use of large language models to adapt message delivery to user context and affective state \cite{jacobson2001}. Together, this body of work highlights the potential of brief, context-sensitive, and socially informed message-based interventions—as well as conversational techniques—as promising strategies for driving behavior change through accessible and lightweight digital health tools. 
\subsection{Engagement and Barriers in Intervention Design}
Engagement in digital health interventions is a complex and multidimensional construct, often described through both subjective experience and observable behavior. Subjective engagement refers to users’ cognitive and emotional states—such as focus, motivation, enjoyment, and immersion—during interactions with an intervention \cite{baron1986, Csikszentmihalyi1990, ObrienToms2008}. Meanwhile, behavioral engagement captures concrete patterns of use, such as frequency, duration, depth of interaction, and adherence over time \cite{McClure2013, Danaher2006}. These dimensions are often used together to assess both the intensity and sustainability of user involvement. Frameworks like the TWente Engagement with Ehealth Technologies Scale (TWEETS) \cite{kelders2020psychometric} offer structured ways to measure these constructs beyond basic usability metrics.

Despite increasing sophistication in measuring engagement, a persistent challenge across digital health literature is the difficulty of sustaining it. Prior studies report steep drop-off rates with digital health tools: Waller and Gilbody \cite{Waller2009} found that clients were twice as likely to discontinue use compared to traditional therapy, and Baumel et al. \cite{Baumel2019} observed median retention rates of only 3.9\% after 15 days. These patterns underscore the need for designs that fit seamlessly into users’ routines while remaining personally meaningful and minimally disruptive.

Several known engagement barriers include repetitive content, poor personalization, and mistimed prompts \cite{Bhattacharjee2023, Brown2014, RennickEgglestone2016}. Interventions that rely on generic or impersonal messages may fail to resonate with users’ evolving needs and situational constraints \cite{Bhattacharjee2022-2, Slovak2023}. Furthermore, digital health tools compete for attention in saturated digital environments, where even personalized content can be ignored if not delivered at the right moment \cite{rolls2014, Muench2017}. To address these issues, researchers have explored methods such as tailoring message timing through JITAIs \cite{deci2020, kabat1994}, improving interface usability \cite{Lipschitz2023}, and embedding interventions within users’ existing communication platforms \cite{Bhattacharjee2022}. However, even with personalization, barriers like life disruptions, social roles, and time scarcity remain difficult to predict or adapt to in real time \cite{Ritterband2009}. These findings suggest that engagement is not just about message design—it’s shaped by complex, external forces such as daily routines, resource access, and emotional readiness.

Building on this body of work, our study investigates engagement in the context of ultra-short interventions. We operationalize engagement using task completion rates, video click-throughs, and the depth of responses to follow-up questions. Given the brevity of each intervention, we emphasize usage frequency and adherence as core engagement metrics, which reflect the momentary and recurring nature of micro-interventions. 

\section{ Study 1: Formative Design Probe}
We first conducted a formative Wizard-of-Oz (WoZ) study to explore how individuals engage with ultra-short, one-minute healthy habit interventions in their daily routines. Rather than testing predefined hypotheses, this study aimed to surface design insights about how prompt characteristics—such as clarity, timing, and actionability—influence engagement. 

\subsection{The Interventions Design Space}
Prompts were designed based on two primary principles. First, \textbf{actionability:} each prompt encouraged a simple, specific action that users could complete quickly (one minute), minimizing cognitive effort and increasing the likelihood of immediate execution. For instance, prompts like “Take a deep breath and hold it for five seconds” required minimal resources and clearly specified the action, increasing the likelihood of success. Actionability was further reinforced by linking actions to immediate benefits, such as “Can’t remember the last time you had water? Your body may need it more than you think. Go ahead, grab a glass or take a few sips right now and notice how it feels.” This approach aligned with Clear’s \cite{clear2018} emphasis on instant gratification and visible progress, motivating users to follow through. Additionally, users were given opportunities to track their task completion through simple confirmations, like clicking a link or replying to the prompt. This added layer of accountability strengthened habit reinforcement and engagement. Second, \textbf{guided engagement}: each prompt included a link to a short instructional video that demonstrated the suggested action. These videos served both as low-cognitive-load guidance and as a measurable engagement marker, enabling us to track click-through rates as a proxy for user interaction with the intervention.

Activities were selected across four categories— Eating and Hydration, Exercise, Mental Well-being, and Mindful Screen Time— based on their feasibility for one-minute completion and relevance to daily routines. While these domains are diverse, they reflect common areas targeted by behavior change interventions and allowed us to explore how ultra-brief prompts function across varied types of motivation, effort, and context. For example, nutrition-focused prompts like “Swap a processed snack for a whole food,” “Plan your next meal or snack,” or “Remove one unhealthy item from your plate” encouraged mindful, accessible food decisions that required minimal effort while supporting long-term health behaviors \cite{Klasnja2009}. Exercise prompts such as “Quick yoga stretch,” “March in place,” or “Desk push-ups” promoted circulation and reduced sedentary time through achievable movement bursts \cite{Consolvo2006, Epstein2016}. Mental well-being activities like “Box breathing,” “60-second mood check-in,” or “Gratitude reflection” aimed to foster emotional awareness and self-regulation \cite{Kauer2012, Sanches2010}. Prompts under mindful screen use—such as “Micro-meditation,” “Screen use intent check,” and “Posture and light check”—targeted digital fatigue and encouraged intentional device use \cite{Hiniker2016, Paredes2018}. These one-minute actions were informed by prior research in HCI on micro-behaviors and in-situ interventions that fit into users’ daily lives.

\subsection{Participants}
Our participants consisted of 11 people (6 women and 4 men, and one participant who preferred not to disclose their gender) aged 18 and above, from diverse backgrounds. Participants' ages ranged from 18-54,  with most falling between 25 and 34 years old (n = 6), followed by 35–44 (n = 2), 45–54 (n = 2), and one participant aged 18–24. Regarding employment status, five participants were employed full-time, two were employed part-time, and two were students. Educational backgrounds varied, with participants reporting qualifications ranging from some post-secondary education to advanced degrees, including Master’s and PhDs. We recruited them through social media platforms, including LinkedIn, to ensure a diverse pool of participants, ensuring the study captured insights from individuals with varied life experiences and routines. Seven participants signed up for a follow up interview as well.
\subsection{Procedures}
The research protocol received approval from the Research Ethics Board at the institution of the first author. Interested individuals first completed an onboarding survey where they provided informed consent, shared their phone numbers, indicated their preferences regarding the four habit categories (healthy eating, mindful screen time, mental well-being, and exercise), outlined their personal goals, and selected their preferred times for receiving interventions.

We used WhatsApp as the delivery platform. Prompts were sent manually by the research team using a WoZ approach to simulate a rule-based chatbot. Participants could also initiate an intervention at their convenience by typing “Ready” into the WhatsApp chat, triggering the next available prompt. This participant-initiated mechanism supported flexible engagement within the study framework.

Over the course of the study, each participant received between 4 to 6 prompts, depending on how many daily time slots they made available. Those who provided only one time slot received four interventions , while those who provided multiple time slots received six during four days of the study. This decision was made by the research team to optimize delivery timing based on participant availability, ensuring interventions arrived during potentially receptive moments without overwhelming users. Researchers tracked whether participants completed the tasks, their click rates on the shared video links. To understand participants’ reasoning behind their behavior, we implemented a structured follow-up flow. When participants confirmed they completed a task (e.g., by replying “Done”), the researcher followed up with a question about why they chose to do it. If no response was received within two hours, a gentle follow-up was sent. If there was still no response, the researcher asked about barriers to completing the task. This approach helped surface both motivators and obstacles to micro-intervention engagement.

After participants had experienced 4-6 interventions, they were invited to a follow-up interview. Each interview lasted around 30 minutes, during which participants provided feedback on the interventions and shared insights into their experiences. Participants were compensated \$20 for their time. At the start of each interview, the researcher explained the study’s purpose: understanding how personalized prompts can encourage one-minute healthy habits. Participants were informed about the session structure, which included questions about their experiences with the interventions, the messages’ design and effectiveness, and suggestions for improvement. With consent, the session was recorded, and notes were taken, with all responses anonymized. The interview covered participants’ initial reactions, the clarity and simplicity of the messages, their relevance and usefulness, and whether the messages motivated them to act. Participants were also asked about barriers to action, how they would personalize the messages, and their suggestions for improving the design. Finally, participants provided feedback on the overall effectiveness of one-minute activities in fostering behavior change and shared additional ideas for interventions.

\subsection{Findings from study 1}
This section explores participants' engagement with prompts, the effectiveness of follow-up questions, and the factors that influenced their responses. Our data consisted of qualitative feedback from interview transcripts, observation notes, and logs of participants' interactions with the interventions. The transcripts were generated from audio recordings of the interviews using Otter.ai, and interaction data included task completion rates and click rates on video links shared during the interventions. For qualitative data, we coded and analyzed this data using reflexive thematic analysis [25], guided by our research questions and design principles. Our pre-existing codes were based on the core elements of actionability, engagement and perceived benefits, which shaped our understanding of participants' interactions and feedback on the interventions.
\subsubsection{Actionability}
The actionability principle, aimed at prompting specific, achievable actions within one minute, showed mixed success. Many participants responded positively to the short duration of tasks (eight participants out of eleven), noting that it made the activities manageable and less likely to be postponed. This suggests that for some users, simple, low-effort prompts were particularly effective in triggering immediate action — aligning with what we later conceptualize as the Immediate Action profile. For example, one participant expressed, “The short duration made the prompts easy to follow” (P7), while another described the prompts as “simple tasks that made me feel good and broke up the monotony in my day” (P10). 
However, other participants faced barriers not just in time, but in cognitive or emotional alignment with the prompt. Tasks that required preparation (like grabbing water or preparing a snack) or were misaligned with personal goals introduced friction that participants could not easily overcome, hinting at the importance of reflective, personalized prompts — a pattern that would later inform our Reflection-First flow. For instance, one participant remarked, “I didn’t feel like it was adding value to my life” (P5), suggesting that beyond simplicity, the relevance and perceived meaning of the prompt was critical.

Participants generally responded positively to prompts. However, non-responses often correlated with barriers such as time constraints or lack of clarity on the effectiveness of the task. Video links achieved moderate-to-high click rates, with participants with high and medium engagement level frequently acknowledging the usefulness of guided instructions. For example one participant (P8) noted, “It was easy to complete” after viewing the video. While other participants felt the videos could be more advanced and engaging in content, they acknowledged finding them useful overall. Participants’ initial reactions to the prompts varied, revealing a divide in how the messages were perceived. Some participants appreciated the personal touch of the prompts (“They sounded very human and had a personal touch” P6), while others found them impersonal and generic (“It felt very automated” P5). Most participants with high engagement levels viewed them as an opportunity to build healthy habits.

Participants encountered barriers to action both before and after attempting to engage with prompts. \textbf{Pre-action barriers} included actions that required preparation, such as preparing snacks or grabbing a glass of water, which were particularly challenging compared to those that required no resources. Situational constraints also played a role, as participants cited examples such as being on a ski holiday or spending time with kids (P4, P5). For instance, a prompt encouraging reflection on the next snack was deemed impractical or irrelevant to daily routines. For some, this was because eating a healthy snack was not a priority or something they cared about, while others mentioned lacking ideas about what constituted a healthy snack. One participant specifically noted the difficulty of going out to find or buy a healthy snack, highlighting the logistical challenges involved (P6). \textbf{Post-action barriers} emerged when participants perceived tasks as irrelevant or misaligned with their personal goals. One participant remarked, “I didn’t feel like it was adding value to my life” (P5), reflecting a broader trend where prompts deemed uninteresting or unrelated to individual objectives reduced engagement. In fact, three out of eleven participants reported disengagement with such prompts. While the short duration of tasks was a noted strength, the findings underscore the critical importance of ensuring relevance and alignment with user goals to enhance the actionability of interventions.
\subsubsection{Engagement}
Participants were categorized into high, medium, and low engagement groups based on their response rate to the prompts. Those who responded to 4-6 prompts were classified as highly engaged, participants who responded to 2-3 prompts were labeled as moderately engaged, and those who responded to 0-1 prompts were categorized as low engagement. Out of the 11 total participants, seven fell into the high engagement group, two were moderately engaged, and two demonstrated low engagement. In total the percentage of engagement is approximately 85.7\%, the percentage of participants clicking on the provided video links is approximately 71.4\% ( see more details in Table 1). This high level of engagement may be partially influenced by the Hawthorne effect \cite{McCarney2007}, as participants were aware that their responses were being observed. Engagement and click-through rates were calculated individually for each participant by dividing the number of prompts they reported completing or clicked on by the total number of prompts they received during the study.
\begin{table}
  \caption{Participant Engagement and Click Rate Summary.  Participants are color-coded by engagement level: red for low engagement (< 50\%), yellow for medium engagement (50–75\%), and green for high engagement ($\geq 75\%)$.}
  \label{tab:eng}
  \begin{tabular}{c>{\columncolor{mygreen}}c>{\columncolor{myyellow}}c>{\columncolor{myred}}c>{\columncolor{myred}}c>{\columncolor{myyellow}}c>{\columncolor{mygreen}}c>{\columncolor{mygreen}}c>{\columncolor{mygreen}}c>{\columncolor{mygreen}}c>{\columncolor{mygreen}}c>{\columncolor{mygreen}}c}
    \toprule
    Participant ID&P1&P2&P3&P4&P5&P6&P7&P8&P9&P10&P11\\
    \midrule
    Engagement Rate (\%)& 83.3& 60&0&20&40&100&83.3&83.3&100&75&100\\
    Video Click Rate (\%)& 83.3& 60&0&40&20&100&66.7&83.3&100&75&100\\
  \bottomrule
\end{tabular}
\end{table}
The engagement data revealed a divide between participants who thrived on straightforward, externally driven cues and those who needed more internal reflection and meaning to stay engaged. High-engagement participants often highlighted the usefulness of reminders arriving during predictable moments like work hours, supporting the idea that external triggers combined with simple actions — characteristic of the Immediate Action flow — were highly effective. As one participant put it, “They reminded me to take breaks during work hours” (P6).

By contrast, participants who struggled to engage often cited the need for personalization and alignment with internal goals, a hallmark of what would later be formalized as the Reflection-First flow. One participant noted, “It would have been nice to choose what time in the day to receive the messages” (P9), while others suggested that generic prompts limited engagement. These reflections hint at the importance of self-awareness and autonomy in designing prompts for sustained motivation.

Participants reported positive emotional and physical outcomes after completing tasks, further reinforcing engagement. One participant stated, “I felt great! I had more energy to get up and go shopping” (P1), while another noted, “I feel good” (P1). The open-ended nature of the questions in the interventions provided flexibility and personal relevance, as participants could adapt the task to their circumstances (“The questions were open-ended, so I could think about coursework or something from my boss” P10). Despite these successes, Low engaged participants felt the prompts lacked personalization, reducing their relevance. One participant noted, “It would have been nice to choose what time in the day to receive the messages” (P9). Others mentioned that the generic nature of the prompts limited their engagement, suggesting that more tailored content aligned with their individual goals and preferences would be more effective. For instance, one participant recommended categorizing prompts based on user feedback after an initial trial period to ensure they are more personalized and relevant moving forward (P9). Participants also highlighted opportunities to enhance engagement through features like progress tracking and dynamic reminders. Suggestions included adding a streak feature to acknowledge consistent task completion and sending adaptive reminders based on user behavior. As one participant proposed, “You’ve completed it five days in a row; don’t lose your streak” (P9). This aligns with findings from games user research showing that streak-based feedback can enhance motivation by reinforcing goal salience and creating a sense of progress \cite{Mekler2013}. These recommendations point to the potential of integrating feedback mechanisms and personalization to foster long-term engagement.

Most participants who ignored the interventions did not provide any reason for not completing the tasks. However, those who were frequently engaged and missed one or two activities cited specific barriers, such as time conflicts and situational challenges (P10: “I couldn’t complete that habit because I had a class then and forgot to follow up on it”). Others mentioned logistical constraints, such as not having the necessary resources at the moment. Despite these barriers, some participants completed the activity after a follow-up, noting that it became easier when reminded (P1: “It was easier to do but I couldn’t find time to do it during the day”). This highlights a clear need for strategies to gain more insights from participants who do not complete the tasks, ensuring a better understanding of their barriers and enabling more targeted intervention improvements.
\subsection{Engagement Patterns and Challenges}
\subsubsection{Why the Prompts Worked for Some Participants}
Successful interventions were typically those that were clear, simple, and easily slotted into existing routines, aligning with what we later identify as the Immediate Action style. This was supported by participant feedback describing the prompts as “easy to follow” (P7), “simple tasks that made me feel good and broke up the monotony in my day” (P10), and “not a lot of time, but something you can do” (P9). Eight out of eleven participants emphasized that the short duration and clarity of the prompts made them manageable even on busy days. Participants appreciated the straightforward nature of the tasks and the low effort required to complete them even on busy days: “It was nice to stop, reflect, and take it easy” (P10). They reported positive emotional and physical outcomes, reinforcing that quick actions paired with external prompts can effectively drive short-term engagement. Many participants reported feeling happier or more energized after completing tasks, reinforcing the short-term emotional and physical benefits of the interventions (“I felt great! I had more energy to get up and go shopping,” P1). 
\subsubsection{Why It Didn’t Work for Others}
On the other hand, when prompts felt misaligned, irrelevant, or insufficiently personalized, they failed to engage participants — a challenge that would later motivate the design of the Reflection-First flow. Some participants wanted clearer expectations upfront and expressed the desire for prompts that aligned with their current mood, needs, or deeper goals. For instance, a participant unavailable during a ski holiday emphasized the need for clear expectation-setting, while others proposed categorizing prompts based on individual preferences. These insights collectively highlighted the importance of both fast-action pathways and reflection pathways to accommodate diverse user needs, directly informing the structure of Study 2.
\subsubsection{Emerging Implications for Flow Design}
Taken together, these patterns revealed a critical insight: engagement success was not uniform across all participants or all prompt types. While some users consistently responded well to immediate, low-cognitive-load prompts that emphasized direct action, others were more engaged when prompts allowed space for self-reflection, contextual alignment, and personal meaning before taking action.
Importantly, these differences were not just random variation in individual responses; they reflected recurring patterns of engagement that emerged inductively from our qualitative data in Study 1. We therefore synthesized these findings into two conceptual engagement flows that would directly inform the design of our next phase:
\begin{itemize}
    \item Immediate Action Flow — structured around fast, clear, action-oriented prompts requiring minimal reflection, designed to trigger quick engagement through simplicity and external cues.
    \item Reflection-First Flow — designed to prompt a moment of self-awareness or intentional reflection before action — aims to support deeper engagement, internal motivation, and personalization. This approach may also enable the formation of implementation intentions, a process by which individuals mentally link a situational cue with a specific behavior, increasing the likelihood of follow-through by bridging intention and action \cite{Gollwitzer1999}.
\end{itemize}
These patterns informed our next phase, where we formalized the distinction between two engagement flows and systematically tested how prompt structure might affect real-world engagement.

\section{Study 2: Flow, Fit, and Co-Creation}
Based on the engagement patterns observed in Study 1, we designed two structured prompt flows for Study 2: Immediate Action Flow and Reflection-First Flow. The Immediate Action Flow aimed to facilitate quick behavior execution with minimal cognitive effort by providing direct, specific action instructions. In contrast, the Reflection-First Flow encouraged users to engage in brief self-awareness or contextual reflection before completing the suggested action, supporting greater intentionality and personalization. Each flow was developed to maintain the one-minute action constraint, while varying the cognitive demands placed on participants during engagement. We also incorporate a co-creation component, inviting participants to rewrite messages to better suit their tone, intent, and daily context. This approach was not only participatory, but also diagnostic: it allowed us to examine how participants interpreted, adapted, or enhanced prompts to make them more engaging. By analyzing these rewrites, we were able to surface deeper insights into how prompt clarity, emotional resonance, and perceived usefulness influence engagement — complementing our behavioral and qualitative findings. In doing so, we moved beyond asking which prompts worked, toward understanding why, when, and for whom they felt worth doing. This phase of study also received approval from the Research Ethics Board at the institution of the first author. 

\subsection{Procedure Of Study 2}
We recruited participants using Prolific, a platform designed for remote behavioral studies, aiming for geographic and lifestyle diversity. All participants were required to have access to WhatsApp and completed an onboarding survey, which includes providing informed consent, demographic details, behavioral preferences. Once enrolled, participants receive a welcome message on WhatsApp, providing instructions on how to interact with the interventions and reinforcing their initial commitment. Habit interventions are delivered through WhatsApp using a structured message flow designed to optimize engagement (the Dialogue flow has been illustrated in Appendix \ref{appendix:a}).

We employed a within-subjects design over a 14-day period. Each participant received one prompt flow—either action-based or reflection-based—randomly assigned for the first 7 days, followed by the other flow during the second 7 days. This within-subjects crossover design allowed each participant to experience both intervention types in a counterbalanced order. Prompts were delivered via WhatsApp once per day. Each prompt was drawn from one of four habit categories—physical activity, mental well-being, healthy eating, and mindful screen use—randomly assigned each day so that participants experienced all categories over the study period.

Participants located across four different time zones (EST, CST, MST, PST). Rather than allowing users to select delivery times, prompts were sent during commonly shared optimal windows for all time zones: morning (10 AM–12 PM EST), afternoon (3–6 PM EST), and evening (7–10 PM EST). Messages included a 1-minute activity suggestion and a link to a short video demonstration. Participants were encouraged to complete the activity and respond with “Done.” If no response was received within two hours, a gentle follow-up was sent, asking whether they had a chance to try the task. Follow-ups also included mood check-ins and reflective questions to deepen engagement. If participants repeatedly missed prompts (after 3 consecutive days), a check-in message was sent offering them the option to continue or exit the study. Participants were asked to engage in rewriting prompts on days 3, 5, 9, and 14. Involving participants in rewriting prompts served as a lightweight co-design method that offered several benefits. First the rewritten messages act as a generative source of  new content ideas that may resonate more deeply with users. Second the process reveals users preferences, motivations and interpretations of prompts, providing insights into what type of messaging feels relevant, actionable or supportive. Finally, prompt rewriting is an accessible strategy for participatory design enabling participants to shape interventions with their experience and values.

All participants who engaged with the prompts at least once,were invited to take part in a follow-up semi-structured interview conducted via Google meet. Interviews were scheduled within one week of the intervention ending. Each session lasted approximately 30–45 minutes, and participants were compensated \$20 USD for their time.The goal of the interview was to understand participants’ subjective experiences with different types of habit prompts, including which messages were most impactful, how motivation fluctuated throughout the study, and how message design influenced engagement. Examples of interview questions included:
\begin{itemize}
    \item What was the most helpful or memorable prompt you received during the 14-day period? Why did it stand out to you?
    \item Were there any moments when you felt less motivated to follow the prompt? Can you describe what was going on at the time?
    \item Did you find any difference between the types of messages? Which format worked better for you, and why?
    \item Can you recall any instance when the prompt led you to reflect more deeply on your habits or well-being?
    \item Were there any specific days or message types that felt harder to engage with? Why?
\end{itemize}

\subsection{Participants}
We recruited 28 participants aged 18 and above, residents of North America, from diverse educational and professional backgrounds.While not statistically powered, this sample size aligns with prior exploratory studies in HCI and CSCW involving longitudinal, in-situ interventions [e.g., 6,123], and enabled both within-subject comparisons and qualitative analysis over the 14-day period.Participants identified as 16 women and 12 men. The age distribution spanned from 18 to 64 years, with the majority between 25–34 years (n = 12), followed by 35–44 (n = 7) and 45–54 (n = 5). Educational attainment ranged from high school diplomas to advanced degrees: 12 participants held a Bachelor’s degree, 5 held a Master’s, and 1 held a PhD, others held college or general diploma. Regarding employment status, 14 participants were employed full-time, 6 part-time, 4 were unemployed and seeking work, 3 were unemployed and not seeking work, and 1 was retired. Twelve participants consented to participate in a follow-up interview.
\subsection{Prompt Flows}
The two flows were designed based on complementary behavioral theories:
\begin{itemize}
    \item Immediate Action Flow: Grounded in Fogg’s Behavior Model (FBM), this flow was designed to trigger action when motivation, ability, and a prompt align at the same moment [1,2]. The design lowers hesitation and cognitive friction by offering simple, easily executable directives. The short video and follow-up reinforcement messages further increase motivation and ability by offering encouragement and demonstrating exactly what to do [6].
    \item Reflection First Flow: In contrast, this flow was rooted in Self-Determination Theory (SDT), which emphasizes autonomy, intent formation, and intrinsic motivation. Prompts began with a reflective question to promote awareness [4], followed by a suggestion for action (e.g., “When did you last check your phone—was it intentional?” then a 1-minute reset activity). Follow-up questions aimed to deepen self-understanding and encourage internalization of behavior change. 
\end{itemize}
Both flows included motivational reinforcement and optional follow-up reflection questions to support engagement and accountability [90,91]. Examples of such intervention is in Table \ref{tab:eng}.
\begin{table*}
  \caption{Examples of One-Minute Interventions by Flow and Category}
  \label{tab:eng}
  \begin{tabular}{llp{7cm}}
    \toprule
  \textbf{Habit Category} & \textbf{Intervention Flow} & \textbf{Intervention Example} \\
    \midrule
    Screen Time&Immediate Action& Pause for 1 minute. Look around and name 3 things you can see, 3 things you can hear, and 3 things you can feel. This simple exercise grounds you in the present moment. Watch this quick video to guide you through it: [Video Link]\\ \hline
   Screen time& Reflection-First& When did you last check your phone—was it intentional? Before you open your next app, take one deep breath and try this 60-second focus reset. A quick pause helps break the autopilot habit, clears mental fog, and refreshes your mind. Give it a try! [Video Link] \\ \hline
  Physical Activity &Immediate Action&Take 1 minute for Tree Pose yoga. Click on this video [Video Link] and follow along. It’s a quick and simple way to improve balance, stretch, and reconnect with yourself.\\ \hline
  Physical Activity & Reflection-First & What’s your posture like right now? Are you slouching, tensed up, or sitting comfortably? Take a moment to realign with this quick posture reset—small changes make a big difference! [Video Link] \\ \hline
  Mental Well-being&Immediate Action&Take 1 minute to sit still, close your eyes, and breathe deeply. Watch this short video to guide you: [Video Link]. Research shows that even one minute in meditation is a simple remedy to unease and to restore your calm.\\ \hline
  Mental Well-being& Reflection-First & What is one challenge you’re facing right now? Take a moment to think about it. Now, imagine the best possible outcome—what does that look like? Watch this short video to guide you through optimistic thinking and discover a small step you can take today: [Video Link]. \\ \hline
  Eating Healthy & Immediate Action & Swap sugary drinks for water or tea. Need help choosing a better alternative? Click here for quick sugar-free swaps [Video Link] . Cutting added sugar helps balance energy and avoid crashes!\\\hline
  Eating Healthy & Reflection-First & Can’t remember the last time you had water? Your body may need it more than you think. Go ahead, grab a glass or take a few sips right now and notice how it feels. Give it a try, you might find it’s exactly what you needed!" click on this link to guide you through it [Video Link]\\
  \bottomrule
\end{tabular}
\end{table*}
\subsection{Co-Creation Protocol}
Participants were prompted to rewrite selected messages after completing them. They could change tone, rephrase, or write a new version. Prompts included:
\begin{itemize}
    \item “Was there anything you would change to make this more helpful?”
    \item “Would you write this differently for someone like you?”
    \item “Can you rewrite a message that would better motivate you?”
\end{itemize}
Rewrites provided insight into user preferences and revealed how tone, specificity, emotional framing, and personal relevance affect engagement.
\subsection{Data Collection \& Analysis}
This study used a mixed-methods approach combining behavioral tracking with qualitative self-report data. In addition to written responses submitted during the intervention, we conducted post-study semi-structured interviews with 12 participants to further explore experiences, barriers, and motivators.
\subsubsection{Engagement Metrics}
To assess behavioral engagement, we tracked message delivery and response logs, including prompt completion, follow-up reply rates, and video click-through data. These metrics allowed us to examine interaction frequency, response patterns, and task adherence. 
\subsubsection{Qualitative Insights}
All interviews were audio-recorded and transcribed using Otter.ai. We employed an inductive thematic analysis approach \cite{braun2019} to analyze the interview data. Initial open coding was conducted in NVivo 15, focusing on participant-reported engagement drivers, barriers, and interpretations of intervention effectiveness. Codes were developed directly from the data and iteratively refined into broader themes that captured how participants experienced and made sense of the intervention.

To support reliability, a second coder reviewed approximately 20\% of the transcripts, using a cross-checking approach to validate interpretations. Rather than conducting full double-coding, the second coder served as a consistency check by reviewing selected excerpts from 2–3 participants. Percent agreement based on overlapping code applications ranged from 73\% to 88\%, with a mean agreement of 80.3\%.
\section{Findings From Study 2}
\subsection{Engagement Metrics}
Across the 14-day intervention period, 28 participants were enrolled in the study. Among them, 8 participants (29\%) completed prompts on all 14 days, demonstrating consistent adherence throughout. While full completion was limited to a subset, most participants engaged with the intervention initially—only 6 (21\%) exhibited no engagement from the outset. Early dropout, defined as disengagement within the first 7 days, occurred in 3 participants, whereas 1 participant dropped off in the second week.

Engagement was assessed using two key indicators: click-through rate (CTR), which measures interaction with the video, and task completion that was measured as self-reported by the participants. While these offer observable behavioral signals of participation, we recognize that they may not fully capture more nuanced effects of prompt flow, such as internal motivation, emotional resonance, or reflective depth. Our goal in using these metrics was to identify broad patterns of responsiveness across conditions, rather than fine-grained psychological shifts. Out of all prompts sent across participants (290 prompts), video click behavior was recorded as follows: 223 videos in the prompts were clicked by the participants (76\%) and 67 videos were not clicked. Table \ref{tab:engagement-metrics} summarizes these metrics across all interventions, as well as broken down by intervention category and intervention flow.
\begin{table}
  \caption{Engagement Metrics by Category and Intervention Flow calculated as CTR = (Clicked / Total) × 100; Task Completion = (Completed / Total) × 100
}
  \label{tab:engagement-metrics}
  \centering
  \begin{tabular}{@{}llcc@{}}
    \toprule
    \textbf{Level} & \textbf{Category / Flow} & \textbf{CTR} & \textbf{Task Completion} \\
    \midrule
    Overall  & All                    & 76\% & 87.2\% \\
    Category & Physical Activity      & 79.7\% & 90.5\% \\
    Category & Eating Healthy         & 70\% & 87.1\% \\
    Category & Mental Well-being      & 80\% & 86.6\% \\
    Category & Mindful Screen Time    & 77.4\% & 84.5\% \\
    Flow     & Immediate action         & 79.5\% & 89.7\% \\
    Flow     & Reflection-First       & 74.5\% & 84.9\% \\
    \bottomrule
  \end{tabular}
\end{table}
Drawing on Fogg’s Behavior Model \cite{fogg2009}, we hypothesized that Immediate Action prompts (Flow 1), by lowering cognitive demands and leveraging fast, automatic responses, would result in higher short-term engagement compared to Reflection-First prompts (Flow 2). This expectation follows the model’s emphasis on reducing friction to facilitate behavior. However, consistent with the findings from Study 1, we also anticipated substantial inter-individual variability, as preferences for autonomy, personalization, and reflective engagement likely moderate the effectiveness of each flow. Additionally, from a motivational perspective, we recognized that Reflection-First prompts may offer distinct advantages over time — by encouraging internalization, self-awareness, and potentially supporting the formation of implementation intentions \cite{Gollwitzer1999}. 

To test this, we compared completion rates across flows using a chi-square test of independence. Completion rates were 89.8\% (123/137) for Flow 1 and 85.0\% (130/153) for Flow 2. The difference was not statistically significant, $\chi^2(1, N = 290) = 1.10,\ p = 0.29$. While our study did not detect significant differences in completion rates or click-through behavior between the two flows, this null result is itself suggestive: reflection-based prompts did not deter participation. In contexts where timing and cognitive readiness align, users may “just as well reflect” — opening the door to deeper engagement without sacrificing accessibility.

Notably, qualitative interview findings revealed that most participants did not explicitly recognize the difference between the two interaction flows during their experience — a result worth highlighting, as it suggests that perceived engagement may be less tied to flow structure and more influenced by personal factors such as timing, mood, or prompt relevance.
\subsection{Qualitative Insights}
In addition to quantitative engagement metrics, qualitative feedback was analyzed to understand participant experiences. The analysis of interviews, and participant written responses during Whatsapp study revealed themes that characterize how individuals engage with ultra-brief (one-minute) behavioral prompts. Below, we present five overarching themes that emerged across the study. Each theme captures not only behavioral responses, but how participants made sense of the prompts — emotionally, cognitively, and socially — over time.
\subsubsection{Situated Simplicity: What Makes One-Minute Prompts Feel Worth Doing}
Although many participants appreciated the brevity and clarity of the prompts, across 12 participants we found that brevity alone wasn’t enough to drive engagement. While many appreciated the one-minute simplicity, they chose to act only when the prompt fit their immediate mood, context, and capacity. This aligns closely with the Fogg Behavior Model \cite{fogg2009}, which posits that behavior happens when motivation, ability, and a prompt converge at the right moment. Participants’ decisions reflected dynamic fluctuations in these elements: they might have been motivated, but if they lacked the mental energy (ability), or if the timing was poor, they skipped the prompt.
For example, P25 noted:“I didn’t have the mental space for that breathing one. But I still drank water.”

This also connects to Bandura’s Self-Efficacy Theory \cite{Bandura1997}, which argues that people engage in actions they believe they can successfully complete. Even if a task was short, participants avoided it if they felt incapable of engaging meaningfully at that moment (e.g., too tired for reflection).
Additionally, participants highlighted that small external barriers — like needing to open a video link or wash a fruit — disrupted the ability component, leading to behavioral shortcuts.
“I didn’t want to wash the pepper, so I had cookies.” (P8)

In contrast, small positive triggers such as a notification badge, a message reminder, or an icon on the phone sometimes acted not just as a nudge but as a satisfying micro “check-off” moment, similar to ticking off an item on a to-do list. For some participants, completing the small task was less about the health behavior itself and more about the emotional reward of completing something that had been sitting on their mental list — a psychological boost tied to progress and closure.
“The icon kept staring at me, so I eventually did it.” (P5)

This reflects an underexplored intersection with task completion theories and the psychology of the Zeigarnik Effect \cite{Zeigarnik1927}, which suggests that people experience a sense of tension when tasks are left incomplete and relief when they are closed. Even micro-prompts, when perceived as tiny commitments, can tap into this motivational pull, providing a small sense of achievement that drives action.

These one-minute prompts were not just quick tasks. For some people, they helped change the direction of their day. One participant said the screen time message helped them reduce their phone use by three hours (P3). Others said the prompts made them realize something about themselves in the moment, such as that they hadn’t eaten or had been holding their breath. Even though the tasks were short, participants felt they were helpful because the effect came quickly. They described feeling more relaxed, more focused, or more motivated afterward. The length was not what made the action useful — the feeling of benefit is what made it valuable.

While momentary fit played a role, a surprising number of participants completed prompts even when the timing felt off—driven by an internal sense that the behavior was “generally good for them.” This suggests a motivational override effect: identity and internalized values sometimes outweighed situational alignment. This aligns with Identity-Based Motivation Theory \cite{Oyserman2009}, which posits that actions are more likely when they affirm a person’s desired self-image, and with Self-Determination Theory \cite{ryan2000}, which explains how internalized motivations can sustain action even when context does not fully support it.
“Even if I didn’t love the timing, I still tried to do it because it was good for me.” (P22)
\subsection{Emotional tone and sensory cues matter more than words}
While one-minute prompts were designed to be clear and concise, eight participants emphasized that what stayed with them wasn’t the wording—but the feeling the prompt conveyed. Participants consistently described sensory elements, emotional tone, and delivery style as more impactful than the specific instructions. In moments of fatigue or stress, the prompts they engaged with most were those that felt gentle, familiar, or soothing—regardless of what they asked them to do.

This aligns with Cognitive Load Theory \cite{Sweller1988}, which suggests that under mental strain, people prefer low-effort, multimodal inputs—such as visuals, sound, or metaphor—over text-heavy or abstract content. As P17 explained:
“The ones with sounds or visuals were easier — I didn’t have to figure anything out.”

In particular, video prompts were often remembered for their emotional tone or calming visuals, even when participants couldn’t recall what the task was. As P28 described:
“I don’t remember the words, but I can still picture that ocean one. It calmed me instantly.”

This memory pattern reflects Dual-Coding Theory \cite{Paivio1986}, which shows that pairing visual and verbal information supports longer-lasting recall. It also connects to mood-congruent memory effects \cite{Forgas1995}, where emotionally resonant cues are more likely to be retained and recalled later.
Tone played a similarly important role. Prompts that used warm, conversational language—like asking a question or offering a suggestion—were described as more motivating than those that were overly directive or instructional. P3 shared:
“The ones that said ‘What made you smile?’ worked better than the ones that told me what to do.”
P9 echoed this sentiment:
“It felt like a friend suggesting something, not an app barking orders.”

Together, these insights suggest that effective micro-interventions don’t just deliver a task—they set a tone. Participants responded best when prompts combined emotional warmth, relatable framing, and sensory anchoring. Rather than prioritizing pure clarity or instructional precision, these findings suggest that emotional tone, delivery format, and perceived gentleness are central to user receptivity—especially in moments of low energy or overwhelm. For ultra-brief interventions, multimodal design may be not just helpful, but essential.
\subsubsection{Ownership and personalization drive sustained motivation}
Nine participants emphasized that feeling a sense of ownership over the prompts—and seeing themselves reflected in the content—was essential to staying motivated. Rather than passively following instructions, they wanted to adapt prompts to their own goals, routines, and values. This desire aligns closely with Self-Determination Theory (SDT) \cite{ryan2000}, which holds that intrinsic motivation is more likely to emerge and persist when people experience autonomy (choice), competence (growth), and relatedness (emotional resonance). As P25 explained:
“If it reminded me why I started — my own reason — I’d be more likely to keep going.”

Participants also expressed interest in scaffolded progression, where small, simple actions could evolve into more meaningful habits over time—allowing for a sense of mastery and development:
“It would be cool if it started with just breathing, then added more later.” (P9)
These preferences echo Habit Formation Theory \cite{Lally2010}, which emphasizes that identity-consistent behaviors become automatic through repetition—but only when the behaviors feel relevant and rewarding. Participants also described a desire for implementation intentions—structured "if–then" reminders linked to personal goals—that could reinforce intention and accountability.

Interestingly, some participants used the prompts not only to act, but to explore who they wanted to become. As P9 noted:
“I tried all the prompts to see which ones would become part of my routine.”
This reflects Possible Selves Theory \cite{Markus1986}, where individuals use behavior to test out future-oriented self-concepts, and Self-Concept Formation theory \cite{Oyserman2009}, which views identity as shaped and refined through everyday action.

Together, these findings suggest that even one-minute prompts can support deeper engagement when users feel in control, see progress, and recognize themselves in the message. When prompts become tools for identity shaping—not just habit triggering—they offer more than momentary nudges; they become part of a self-directed process of growth.
\subsection{Findings From Participants Co-Creation Activity}
When participants were invited to rewrite selected prompts, their edits revealed not just surface preferences, but deeper efforts to shape the tone, purpose, and delivery of behavior change support. These rewrites reflected a shift from passive receipt to co-authorship—transforming prompts into personal scripts that fit users’ own motivations, emotions, and daily rhythms. This aligns with Self-Determination Theory’s \cite{DeciRyan2000, RyanDeci2017} emphasis on autonomy and with HCI work on participatory design and design appropriation, where users adapt content to reflect their identity, goals, and values \cite{Rosner2014, Bodker2004}.
\subsubsection{Rewriting for Tone, Guidance, and Emotional Fit}
Participants favored prompts that felt supportive and low-effort, especially for movement or mindfulness activities. Instead of standalone instructions, they preferred guided, conversational delivery that reduced cognitive demand:
“Slowing it down and making it more of a guide instead of an instruction… something you could follow along with instead of instructions you need to remember.” (P19)
Rewrites often added a friendly or inclusive tone (“Hey y’all! How ya feeling?” – P8), or emphasized ambient and sensory context (“A video filmed outdoors might be nice, with birds chirping.” – P28). These edits suggest that tone, pacing, and environmental resonance mattered as much as the behavior itself \cite{Orji2019}.
\subsubsection{Rewriting for Immediate Relevance and Motivation}
Other participants modified prompts to include short-term benefits or emotion-linked rationale. These changes reflected an effort to connect the task with something that felt personally helpful or needed in the moment:
“Maybe just a note about how it could help so that I can envision what it could help me with.” (P20)
They also emphasized adaptive timing and emotional fit:
“It would be more helpful if it reminded me of how I’m feeling or what I need today.” (P3)
These examples show how participants sought prompts that not only fit behaviorally but also emotionally and contextually—echoing work on affective computing and context-aware interventions \cite{Klasnja2011}.
\subsubsection{Rewriting as Narrative and Identity Work}
A few participants took this further, rewriting prompts into micro-stories or motivational scripts drawn from lived experience. These were often expressive or reflective, such as reframing a hydration reminder with a backstory about energy crashes, or imagining prompts that scaled over time.

These rewrites suggest that prompts can serve as tools for identity expression—not just reminders, but reflections of “the kind of person I want to be.” While infrequent, these instances hint at the potential for micro-interventions to support self-construction over time.
Participants did more than adjust phrasing—they reshaped prompts to feel like extensions of their own voice, logic, and goals. These edits offer early evidence that allowing users to co-author micro-intervention content may enhance not only acceptability but also emotional connection and sustained engagement.
\section{Discussion}
This study contributes to the growing body of research on micro-interventions by examining how individuals engage with one-minute healthy habit prompts in everyday life. We identified design principles that help such prompts feel worth doing — and explored how these design choices play out across different contexts and flow structures. Our discussion addresses the guiding research questions and situates our findings within existing HCI and behavioral literature on just-in-time interventions, momentary engagement, and emotional readiness.
\subsection{RQ1: How can we encourage individuals to engage in one-minute behavioral change interventions?}
Across both written and verbal data, participants described one-minute tasks as appealing only when they felt useful, relevant, and matched to the moment. Brevity was not enough. Participants were more likely to engage when the task addressed a present need—such as physical discomfort, emotional stress, or a feeling of fatigue. These findings provide further evidence for prior work on just-in-time interventions, which shows that actions are most likely to occur when the prompt feels contextually appropriate and cognitively light \cite{Grimes2010, Spring2013}.
“I didn’t do the reflection prompt — I wasn’t in the headspace. But I did the water one. That was easy.” (P25)

Consistent with cognitive load theory and real-world engagement studies \cite{Hayes2008, Hekler2016}, we found that physical prompts (e.g., stretching, sipping water) were easier to complete than cognitive or emotional ones (e.g., gratitude reflection, emotional check-ins). Participants emphasized that even small decisions required attention, and they appreciated prompts that respected low-energy states by offering “something doable even when tired.”
Moreover, many participants mentioned completing prompts later—sometimes hours after the initial delivery. This delayed completion suggests that even micro-prompts can have lingering effects on attention, pointing to a need for more flexible definitions of engagement beyond immediate response \cite{Pina2015}.

Participants also reported that being asked a follow-up question increased their likelihood of completing the action. Several framed it as a form of soft accountability or self-check-in—making them feel seen or prompting a gentle pause for evaluation. These low-stakes, conversational touches often served as reflection scaffolds that not only encouraged behavior completion but also deepened emotional resonance. This aligns with prior work on conversational health interventions, where follow-up questions and micro-dialogue enhance both compliance and perceived support. These responses also suggest a link to meta-cognitive engagement \cite{Flavell1979}: prompts that invite users to reflect on their state or intention may activate critical awareness of their behavior in context—what they’re doing, why it matters, and whether it aligns with their broader goals.

While many prompts were successful when they matched users’ current state, we also observed cases where participants engaged despite poor fit—driven less by situational alignment and more by internalized values. Rather than relying solely on environmental triggers, some users treated prompts as affirmations of their broader health goals. This suggests that micro-interventions may draw strength not only from momentary readiness but also from identity-congruent motivation, offering opportunities to reinforce behaviors even when conditions are suboptimal.
\subsection{RQ2: How can concrete examples be crafted to encourage engagement?}
Prompts were most effective when they were specific, emotionally warm, and visually or sensorily grounded. Participants preferred prompts that clearly described the action, explained why it might help, and included simple affirmations or encouragements.
“Even a small ‘this helps your focus’ line made it easier to try.” (P28)

These findings provide empirical context for persuasive design and health behavior research that highlights the value of clear affordances and brief rationales \cite{Zimmerman2007}. Participants responded well to prompts that included visual imagery, simple metaphors, or optional videos—especially when these reinforced the physicality of the action (e.g., a stretch, a sip of water, a pause to observe the environment). These sensory-oriented strategies reduced ambiguity and made the behavior feel concrete and achievable, particularly under conditions of stress or limited attention.

Participants also valued prompts that felt emotionally supportive rather than directive. The tone of the message—whether it felt friendly, encouraging, or human—had a strong impact on how the prompt was received. Prior studies on affective interaction and self-compassion technologies emphasize that tone can shape both compliance and perceived value \cite{Hayes2011, Murnane2015}, and the current findings show when and how this occurs.

Beyond tone and clarity, what stood out was that emotional resonance and affective imagery often left a deeper impression than the specific behavioral instruction. Participants retained the feel of prompts—calming, energizing, or grounding—even when the exact words or actions faded. This pattern suggests that short interventions may function more like emotional anchors than task-oriented reminders, pointing to an underutilized role for sensory and affective design in shaping memory and meaning.

In later stages of the study, participants showed growing confidence in modifying or substituting prompts to better suit their routines. This suggests that effective examples may not only support initial adoption, but may also scaffold adaptive internalization of the behavior \cite{Hamari2015, Morgan2012}.
\subsection{RQ3: How can different interaction flows impact engagement?}
We tested two interaction flow types — Immediate Action (directive prompts to act immediately) and Reflection-First (prompts inviting participants to pause and reflect before acting) — to explore how structural design influenced engagement. Our findings reveal a nuanced picture: participants varied in which flow they found easier or harder, depending largely on their momentary state.

Some participants reported that shifting from their current task into a reflection prompt was difficult, especially during busy or mentally demanding moments:
“I didn’t have the mental space for that breathing one. But I still drank water.” (P25)

Others, in contrast, found activity-oriented prompts harder, especially when they were physically tired or in constrained environments. For example, one participant said that mindfulness prompts were more accessible in the moment, while physical prompts like stair climbing were easier to skip.
 “Switching from thought to body is hard for me — I just wasn’t in that headspace.” (P28)
 
Some participants appreciated both flow types and saw value in the diversity of tasks and prompts. Importantly, neither flow type inherently blocked participants from acting. Most described completing prompts not because of the flow structure but because they saw them as generally useful for their health and well-being — unless the timing clashed sharply with their immediate situation.

This highlights that momentary fit — whether the person had mental, emotional, or physical capacity for the prompt — was the more decisive factor, consistent with the Fogg Behavior Model \cite{fogg2009} and Self-Determination Theory \cite{deci2020}. What mattered most was whether the prompt aligned with participants’ motivation, ability, and sense of personal relevance at the time.

Click-through data further showed that embedded video prompts improved early follow-through but were often skipped after initial exposure, reinforcing the idea that media elements serve as early scaffolds rather than ongoing engagement drivers \cite{Mankoff2013}.

Follow-up questions sometimes nudged participants into engagement or reflection, even when the original prompt had been skipped:
“I wasn’t going to do it, but then I thought why not.” (P26)

Importantly, we observed that participants’ strongest preferences clustered more around the prompt category (mental well-being, mindful screen time, physical activity, or healthy eating) than flow structure. Participants engaged more consistently when the content matched their goals and interests, suggesting that systems should prioritize adaptive content delivery over rigid sequencing — suggesting how to apply just-in-time adaptive intervention (JITAI) principles \cite{Kay2012, fogg2009}.

Overall, our findings suggest that while flow structure (directive vs. reflective) influences perceived ease – depending on context – neither structure nor category inherently blocked engagement. Instead, engagement hinged on momentary fit, perceived usefulness, and personal relevance — reinforcing the importance of designing flexible, user-centered micro-interventions that adapt to the realities of everyday life. While our work builds on behavioral models like FBM to understand how prompts may trigger action, our findings suggest that this framing alone may not fully explain what makes ultra-brief interventions feel engaging or worth doing. In our study, engagement was rarely driven by frictionless prompting or behavioral simplicity alone. Instead, participants responded most when prompts felt emotionally aligned, personally meaningful, or reflective of their current context—dimensions that fall outside the scope of traditional trigger-based models. This observation complements recent critiques of persuasive design [119], which argue that behaviorist models may underrepresent the affective and interpretive aspects of user experience. Our findings point toward a need for micro-intervention design approaches that incorporate not just triggering mechanisms, but also support for user autonomy, emotional connection, and situated interpretation.

These insights contribute to HCI and CSCW research by indicating that lightweight, ultra-brief digital interventions can flexibly support users across diverse mental, emotional, and physical states — not because of rigid flow designs but because of their adaptive alignment with momentary needs and personal motivations. Previous research has emphasized flow structure and prompt sequencing.nHowever, our findings suggest that in real-world contexts, it is the content category, emotional resonance, and perceived health value — not flow structure alone — that most strongly shape engagement. This reframes what matters most in everyday micro-interventions: not when the prompt arrives or how it's sequenced, but what it evokes, how it feels, and whether it aligns with what users care about.

By showing that neither directive nor reflective flows inherently block action, we expand the design space for micro-interventions. Our findings suggest that cooperative behavior change and self-management are best supported through adaptive, user-centered systems—those that respect autonomy, adjust to real-life contexts, and promote well-being without adding cognitive or time burden. These results inform the development of scalable, socially supportive strategies that can meaningfully fit into the fabric of daily routines.

\section{Limitations \& Future Work}
This study offers useful insights into how people interact with one-minute healthy habit prompts in everyday life, but there are several limitations that should be considered. First, the study lasted only two weeks, and participants experienced more than one type of prompt flow during that time. Even though we used counterbalancing, some participants may have been influenced by the order in which they received the flows. For example, trying one flow first could have shaped how they felt about the second, especially as familiarity increased or motivation changed. A longer study or one that compares flows across different groups might help us understand how timing and repetition influence engagement more clearly.

Second, the participant group was small and mostly based in North America. Most were already comfortable using messaging apps and had some interest in health and well-being. This means the findings may be less likely to apply to people in other regions or with different cultural backgrounds. Tone, timing, and what feels supportive can vary by culture, so future work should test these kinds of prompts with more diverse groups, including those who speak different languages or who live in places where health tools are harder to access.

Another limitation relates to how many topics were included in a short time. While some participants enjoyed trying prompts from different categories, others said switching topics each day made it hard to form a routine. Some wanted to stay with one focus area — like food or stress — to build a habit before moving on. Future studies could offer options to let users pick a focus and stick with it for a few days. This might support stronger engagement by reducing decision fatigue and helping people feel more in control of the process.

While this study focused on engagement outcomes such as task completion, prompt interaction frequency, and response depth, we did not directly measure psychological mediators like self-efficacy or intrinsic motivation, which are core to models such as Self-Determination Theory and Social Cognitive Theory. These constructs are known to influence adherence and long-term behavior change and may have helped explain why certain prompt structures (e.g., reflection-first) felt more or less meaningful to different users. Future studies could incorporate brief, repeated measures of motivation, confidence, or perceived autonomy to more directly assess how micro-interventions support the internal drivers of sustained engagement.

Finally, the system was not able to adapt prompts in real time based on how people were feeling, or how much energy they had. Although participants shared strong preferences about prompt type, timing, and tone, the WhatsApp format limited our ability to personalize messages based on daily context. Future versions of this work could include simple check-ins or sensors to adjust prompts — for example, offering light, physical actions when users are tired or stressed, and reflection prompts when they have more time or energy. This kind of adaptation is being explored in other just-in-time health tools, and it may work well for short, low-effort messages like the ones tested here.

Looking ahead, generative AI offers promising tools for making micro-interventions more adaptive and context-sensitive. Large language models (LLMs), in particular, can dynamically tailor message tone, difficulty, and content based on user input, past behavior, and real-time context—including mood, activity, or expressed needs. This capacity to generate responsive, emotionally attuned prompts may be especially well-suited for lightweight interventions like the one-minute tasks tested in our study. Emerging research in just-in-time adaptive interventions (JITAIs) has begun exploring how conversational agents and LLMs can support moment-level personalization by interpreting natural language cues and adjusting interactions accordingly \cite{Haag2025, Kocaballi2020, Maher2023}. Applying these techniques to micro-intervention design could support deeper user alignment while maintaining the simplicity and immediacy that make these prompts scalable and sustainable.

In future research, we hope to explore how micro-interventions can better fit into people’s daily lives, not just through timing or format, but by learning what works best for each person over time. Even small changes — like offering choice, adjusting tone, or supporting habits that matter most — may help more people follow through, especially when life feels busy, unpredictable, or overwhelming.

\section{Conclusion}
These studies suggest that one-minute prompts may serve as a useful entry point into habit formation—particularly when they feel timely, emotionally attuned, and low in effort. Across two exploratory studies, we observed that participants were more likely to engage when prompts aligned with their current mood, offered small but immediate benefits, or used a tone that felt gentle and supportive. Action-oriented prompts appeared helpful during low-energy moments, while reflection-based prompts were associated with more thoughtful and emotionally resonant responses. Co-created rewrites and follow-up questions seemed to increase personal relevance and motivation for some participants. Taken together, these findings point to how small design choices—such as tone, timing, and prompt structure—can influence how people perceive and engage with micro-interventions. This work shows how designing small and easy actions can support healthy behavior, and can also help researchers create technologies that fit better into people's everyday lives.


\bibliographystyle{ACM-Reference-Format}
\bibliography{healthy}

\appendix

\section{Dialogue Flow}
\label{appendix:a}
\begin{figure}[h]
    \centering
    \includegraphics[width=0.45\textwidth, height=0.9 \textheight]{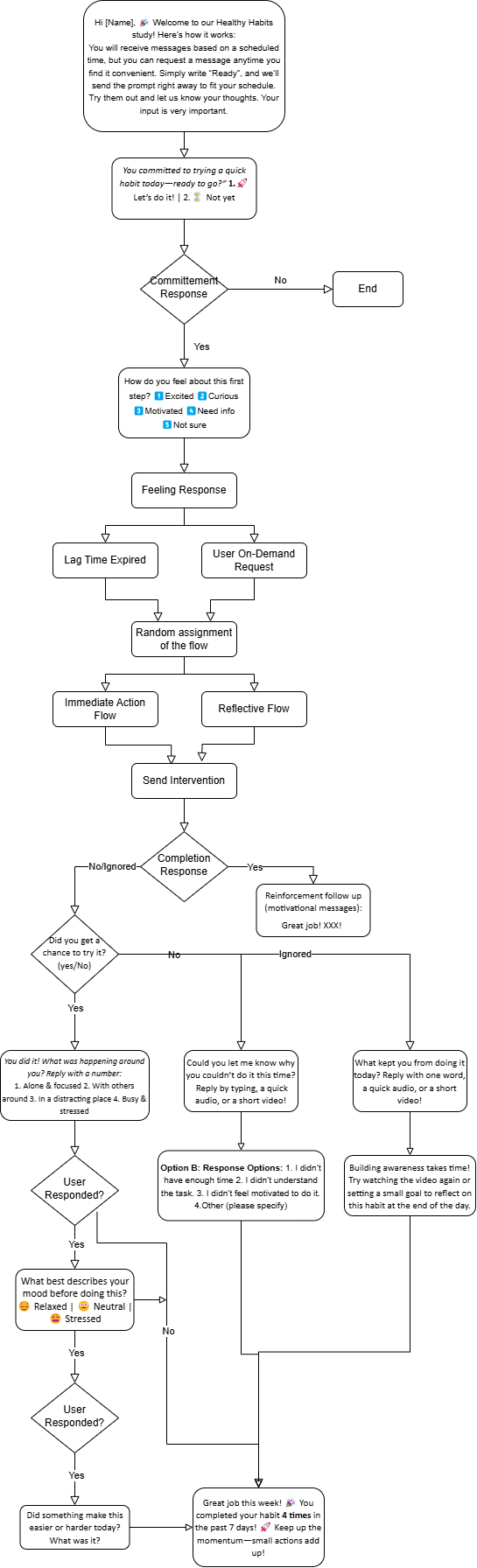}
   \caption{Flow of Participant Interaction in the WhatsApp Study.}
    \label{fig:flowchart}
    \Description{A flowchart showing participant interactions in the WhatsApp study, including initial onboarding, randomized intervention flows, engagement tracking, and follow-ups.}
    \label{fig:flowchart}
\end{figure}

\end{document}